\documentclass[twocolumn,showpacs,amsmath,amssymb,aps,pra,floatfix,nofootinbib]{revtex4}
\usepackage{bm}
\usepackage{amsfonts}
\usepackage{graphicx}
\usepackage{epstopdf}
\begin{document}

\title{Lorentz contraction of the equal-time Bethe-Salpeter amplitude in two-dimensional massless quantum electrodynamics}
\author{Tomasz Rado\.zycki}
\email{t.radozycki@uksw.edu.pl} \affiliation{Faculty of Mathematics and Natural Sciences, College of Sciences,
Cardinal Stefan Wyszy\'nski University, W\'oycickiego 1/3, 01-938 Warsaw, Poland}

\begin{abstract}
The Lorentz transformation properties of the equal-time bound-state Bethe-Salpeter amplitude in the two-dimensional massless quantum electrodynamics (the so called Schwinger Model) are considered. It is shown that while boosting a bound state (a `meson') this amplitude is subject to approximate Lorentz contraction. The effect is exact for large separations of constituent particles (`quarks'), while for small distances the deviation is more significant. For this phenomenon to appear, the {\em full} function, i.e. with the inclusion of all instanton contributions has to be considered. The amplitude in each separate topological sector does not exhibit such properties.
\end{abstract}
\maketitle

\section{Introduction}
\label{sintro}

Lorentz contraction of a rigid body is a well known phenomenon in special theory of relativity. For example the equal time measurement of spacial separation of the two ends of a rigid rod leads to different results depending on the chosen inertial frame. The simultaneity is not an invariant notion in relativistic physics: events that happen simultaneously in different points in one reference frame cease to be such in another frame. This leads to the famous relation between lengths of the rod in question measured in two different frames:
\begin{equation}
l=\frac{l_0}{\gamma},
\label{contra}
\end{equation}
where $l_0$ is the length in the frame related to the rod, $l$ is the length of the rod moving with velocity $v$ parallel to itself and $\gamma=1/\sqrt{1-v^2/c^2}\geq 1$. Hereafter we will omit $c$ and use units, where $c=\hbar=1$. 

Such shortening in the direction of motion should be shared by every body moving relative to an external observer and therefore ought to be found as well in the spacial distribution of quantum probability amplitude, i.e. the wave-function describing a bound state as, for instance, meson. In this case it is not any longer a kinematical effect, but becomes a dynamical one involving fundamental interactions between constituent fermions and consequently it is far from being trivial. This is really the case in collisions of heavy ions, which exhibit strong flattening in the laboratory frame with factor $\gamma$ reaching the value of several thousands~\cite{baltz,singh,wong}. In this frame such colliding nuclei look as thin disks wherefore often called `pancakes'.

Of course we do not expect to find the {\em exact} Lorentz contraction of the wave-function, because -- at least in quantum field theory (QFT) -- it does not contain the whole information about the bound system, which is a complicated structure containing both valence fermions and additional fermion-antifermion pairs, as well as gauge bosons. It cannot be, therefore, described by a single wave-function. Only considering the bound state in its full complexity, one might expect to observe the precise Lorentz contraction for spatial distributions of various physical quantities. But even the `simple' wave-function, spoken of in this paper, should somehow reflect in its behavior the phenomenon known from relativistic kinematics.

Several works have been devoted to the investigation of this effect for the bound-state wave-function in the case of hydrogen atom, positronium, nuclei or model systems~\cite{bro1,bro2,kimza,inozem,kimnoz,hoyer1,gnf,gn,kim,jarv0,hoyer2,diet,hoyer3,simonov}. When the relativistic motion of a primarily non-relativistic system as for instance a hydrogen atom is considered, QFT phenomena appear, and the problem is no longer purely quantum mechanical but requires field theoretical approach. Instead of solving a Schr\"odinger equation for the system in question, one has to cope with Bethe-Salpeter (BS) equation~\cite{bs,gml}. This equation is, however, extremely difficult to be solved even in the simplest model cases. It is a multidimensional integral equation and, to make matters worse, with unknown ingredient functions as full propagators or the interaction kernel, which obviously are unknown in realistic QFT. There are, by now, only few examples, where solutions of BS equation have been found, but unfortunately it has been done for the price of drastic simplifications of the equations and weakly controlled approximations~\cite{wick,cut,nn1}. 

The analysis concerning Lorentz contraction of various bound systems carries then the stigma of this approach. The explicit and accurate verification to what extent this phenomenon is reflected in the behavior of the {\em exact} bound-state wave-functions is until now lacking and, therefore, is worth some attention. The approximated results obtained so far for these systems and model studies confirm the almost-contraction of the wave-functions, although there has also been found a counterexample~\cite{gn}, the so called Bakamjian-Thomas model~\cite{baka}. 

The deviations from the exact Lorentz contraction, that have been found during these investigations, are attributable to the existence of higher sectors, which become important for relativistically moving true bound state, and which are forgotten in the wave-function approach.

It should be stressed however, that there exists a QFT model -- the two-dimensional massless electrodynamics, known as the Schwinger Model (SM)~\cite{js} -- for which the {\em exact} BS amplitude has been found without any simplifications~\cite{trsing,trfactor}  and so far it is probably the unique system with this property. Since its formulation in the early sixties of last century this model has become an excellent testing laboratory for various nonperturbative aspects of QFT. In this context one can mention confinement, topological sectors, instantons and condensates. Other interesting features include the existence of anomaly and nonzero mass generation of the gauge boson without the necessity of introducing any auxiliary Higgs field. Another remarkable property of this model -- particularly important for our present work -- is the existence of a bound state, which might be called a `meson', constituting a system of a fermion and an antifermion (we will call them quarks,  because of the similarity between SM and QCD in some respects as for instance confinement, despite the fact, that we will still talk about `electrodynamics') and which is also known as the Schwinger boson. 

In the following sections we would like to concentrate on the BS amplitude of this bound state and explicitly check, how it behaves under boosts. In Minkowski space this amplitude is a function of two variables (because we are in $2$ dimensions): relative time and relative position of the constituent quarks. To reveal the Lorentz contraction we have to consider the so called equal-time (ET) function, i.e. the function for which the relative time is set to zero. This corresponds to the {\em simultaneous} measurement of the positions of two ends of the rod, in the classical kinematical derivation of this phenomenon. Surely such an ET function in one frame is not a Lorentz-transformed ET function from another frame, because, as mentioned, the simultaneity is a frame dependent notion. Fortunately in the SM we dispose the exact form of the bound-state amplitude in {\em any} frame! It would be a sin not to take advantage of these unique properties of this model to verify thoroughly the phenomenon of Lorentz contraction.

One more point should be emphasized at the end: as mentioned above the SM has a complicated vacuum structure similar to that of QCD. The appearance of the instanton sectors in the theory poses a new question, which, to our knowledge, no one has tried to answer. Namely, how do the components of the wave function in every topological sector transform under boosts? Is each of them subject to the approximate shortening or it concerns only the total amplitude? SM provides the answer to them.

The present paper is organized as follows. In section~\ref{trnsf} we recall the most important properties of the model, section~\ref{bsamp} is devoted to the analysis of the BS amplitude in the center-of-mass (CM) frame and in the frame, where a meson is moving, and in the final section we present numerical calculations in the form of plots, which allow to compare wave-functions in various frames.

\section{Simple description of the model}
\label{trnsf}

In this section we would like to briefly recall the basic properties of the Schwinger Model. It is defined by the two-dimensional Lagrangian density: 
\begin{align} 
{\cal L}(x)=&\overline{\Psi}(x)\bigg(i\gamma^{\mu}\partial_{\mu}-
gA^{\mu}(x)\gamma_{\mu}\bigg)\Psi (x)\nonumber\\
&- \frac{1}{4}F^{\mu\nu}(x)F_{\mu\nu}(x)- \frac{\lambda}{2}\left(\partial_{\mu}A^{\mu}(x)\right)^2, 
\label{lagr} 
\end{align} 
where $g$ is the coupling constant, which bears the dimension of mass, and $\lambda$ the gauge fixing parameter (in what follows we choose Landau gauge setting $\lambda\rightarrow\infty$). It then simply describes the massless electrodynamics in one spatial and one temporal dimension.

The Dirac gamma matrices $\gamma^\mu$'s may be chosen in~(\ref{lagr}) as two-dimensional ones in the form:
\begin{equation}
\gamma^0=\left(\begin{array}{lr}0 & \hspace*{2ex}1 \\ 1 & 0 
\end{array}\right), \;\;\;\;\; 
\gamma^1=\left(\begin{array}{lr} 0 & -1 \\ 1 & 0 
\end{array}\right),\label{g01}
\end{equation}
and $\gamma^5$ is given by
\begin{equation} 
\gamma^5=\gamma^0\gamma^1=\left(\begin{array}{lr} 1 & 0 \\ 0 & -1 \end{array}\right),
\label{g5}
\end{equation}
The metric tensor we use, is defined by the relation $\mathfrak{g}^{00}=-\mathfrak{g}^{11}=1$. 

The gauge boson primarily massless, as required by gauge invariance, acquires a mass~(\ref{mu}) due to the presence of  anomaly~\cite{abh}. This massive particle may be interpreted from the other side as a fermion-antifermion bound state.  

The theory defined by~(\ref{lagr}) has a nontrivial topological structure, which makes it similar to QCD~\cite{cadam1,smil,gattr,maie,rot,gmc}. In the temporal gauge $A_0=0$, the surface of constant time gains the topology of a circle, similarly as one has for the gauge group $U(1)$. This results in the occurence of topological sectors labeled by the winding number of the first homotopy group $\pi_1(S^1)$~\cite{rajar, huang}. There appears the infinite set of vacua -- the so called topological vacua -- which may be denoted by $|N\rangle$, where $N$ is a certain integer number corresponding to the winding index. None of these vacua is the true physical vacuum. 

In the massive version of the model, the phenomena of tunneling between topological vacua occur (described as `instantons'). They lead to the emergence of the physical vacuum, the so called $\theta$-vacuum, in the form of the linear combination (similar to Bloch state in solids~\cite{rajar}):
\begin{equation}
|\theta\rangle=\sum_{N=-\infty}^{\infty}e^{i N\theta}|N\rangle.
\label{tvac}
\end{equation}

Massless fermions, dealt with in the present paper, suppress tunneling, but the $\theta$-vacuum~(\ref{tvac}) still retains its importance, because topological vacua, contrary to $|\theta\rangle$, do not exhibit the cluster decomposition property~\cite{cdg,rajar,cadam1,smil,gattr,rot,trinst}. However, due to the lack of tunneling, while calculating the vacuum expectation values $\langle\theta|\hat{O}|\theta\rangle$, only operators $\hat{O}$ which change $N$ (or chirality, which is equivalent~\cite{rajar,huang}) gain off-diagonal, i.e. instantonic contributions. The example of such an operator is a product of fermion fields $\Psi$, which appears for instance in the propagator:
\begin{equation}
S(x-y)=\langle\theta|T(\Psi(x)\overline{\Psi}(y))|\theta\rangle.
\label{sdef}
\end{equation}
Consequently, due to the $2:1$ correspondence between chirality and topological index $N$, the full propagator has contributions from instanton sectors $0,\pm 1$, which simply means that $\Delta N=0,\pm 1$. It has been found in the explicit form~\cite{trinst}:
\begin{equation}
S(x)=S^{(0)}(x)+S^{(1)}(x),
\label{sf}
\end{equation}
where the first term
\begin{equation}  
S^{(0)}(x)={\cal S}_0(x)\exp\left[-ig^2\beta(x)\right],  
\label{propbet} 
\end{equation} 
was obtained already in the original Schwinger work~\cite{js}. The symbol ${\cal S}_0(x)$ denotes here the free Feynman propagator:
\begin{equation}
{\cal S}_0(x)=-\frac{1}{2\pi}\ \frac{\not\! x}{x^2-i\varepsilon}.
\label{s0}
\end{equation}

The second (or instantonic) term, for $\Delta N=\pm 1$, has the form:
\begin{equation}
S^{(1)}(x)=\frac{ig}{4\pi^{3/2}}e^{-i\theta\gamma^5}e^{\gamma_E 
+ig^2\beta(x)}
\label{s1}
\end{equation}
and the well known auxiliary function $\beta$ appearing in~(\ref{propbet}) or~(\ref{s1}) is defined by  
\begin{align}  
&\beta(x)=\nonumber \\
&\left\{\begin{array}{l}\frac{i}{2g^2}\left[- 
\frac{i\pi}{2}+\gamma_E +\ln\sqrt{ \mu^2x^2/4}+   
\frac{i\pi}{2}H_0^{(1)}(\sqrt{\mu^2x^2})\right], \\
\hspace*{10ex}
x\;\;\;\; {\rm timelike},\\   
\frac{i}{2g^2}\left[\gamma_E+\ln\sqrt{- 
\mu^2x^2/4}+K_0(\sqrt{-\mu^2x^2})  
\right], \\ \hspace*{10ex} x\;\;\;\; {\rm  
spacelike}.\end{array}\right.  
\label{beta}
\end{align} 
$\gamma_E$ is here the Euler constant and $\mu$ is the `Schwinger boson', i.e. bound state, mass:
\begin{equation}
\mu=\frac{g}{\sqrt{\pi}}.
\label{mu}
\end{equation}
Symbols $H_0^{(1)}$ and $K_0$ refer to the Hankel function of the first kind and Basset
function respectively.

Contrary to $S$ the boson propagator
\begin{equation}
D^{\mu\nu}(x-y)=\langle\theta|T(A^\mu(x)A^\nu(y))|\theta\rangle
\label{ddef}
\end{equation}
has contributions only from the trivial sector ($\Delta N=0$) since gauge fields operators have vanishing matrix elements between different topological vacua.

As mentioned in Introduction, SM has a bound state, which may be found in the $t$-channel of the four-point (i.e. two-fermion) Green's function~\cite{eden,mand,trjmn,trsing,trfactor}. The analytical structure in momentum space of the latter reveals the existence of a pole at $P^2=\mu^2$, where $P$ is the total two-momentum of the system. From the residue in this pole the BS function may be read off, without the necessity of solving the BS equation itself. This function will be dealt with in the next section and its form reflects the properties of the fermion propagator~(\ref{sf}), that is, it has contributions from sectors $\Delta N=-1,0,1$. 

The $\theta$-vacuum, constituting the true vacuum of the theory, is invariant with respect to boosts
\begin{equation}
K|\theta\rangle=0,
\label{ktheta}
\end{equation}
where $K$ denotes the boost generator.
This statement does not refer, however, to the particular $N$-vacua, which are not physical vacua and may be Lorentz non-invariant. It was already suggested by Nakanishi~\cite{naka,abe} and manifests itself through the results of the following sections. The (approximate) Lorentz contraction does not hold in each separate instanton sector but is recovered only after the whole function is constructed. 

\section{Bethe-Salpeter amplitude}
\label{bsamp}

The BS amplitude for a bound state may be found either by solving BS equation or by investigating the analytical structure of the momentum-space four-point Green's function to extract the residue in the pole corresponding to a particle in question. Both approaches are extremely difficult to implement not only in a true field theory but even in simple models. As mentioned in the Introduction the Schwinger Model seems to be by now the unique field theory in which the BS amplitude for the formation of a `meson' has been found in the full form, without any approximations.

The well known fact (and also the requirement originating from probability theory) is  the factorization of the residue, which is a product of two BS amplitudes. They can alternatively be found as matrix elements of the kind:
\begin{equation}
\Phi_P(x)=\langle\theta|T(\Psi(x/2)\overline{\Psi}(-x/2)|P\rangle\; ,
\label{bsampl}
\end{equation} 
where $P$ is the two-momentum of the `meson' -- i.e. the bound state -- and $|P\rangle$ denotes the one-boson Fock state state built over vacuum $|\theta\rangle$. The symbol $x=[t,r]$ indicates here the relative variables among constituent `quarks'. As discussed earlier the operator in question is bilinear in fermion fields, and therefore is not diagonal in topological vacuum index $N$. For such an operator -- identically as for fermion propagator -- nonzero matrix elements between different topological vacua (or between Fock states built over them) lead to the appearance of contributions from instanton sectors $\Delta N=0,\pm 1$. In effect the Bethe-Salpeter amplitude $\Phi_P(x)$  may be written as
\begin{equation}
\Phi_P(x)=\Phi_P^{(0)}(x)+\Phi_P^{(1)}(x),
\label{phip}
\end{equation}
where similarly as in~(\ref{sf}) the superscript `$(1)$' refers to both cases: $\Delta N=1$ and $\Delta N=-1$.

The explicit form of $\Phi_P^{(0,1)}$ was given in~\cite{trsing,trfactor}:
\begin{subequations}
\label{phipa}
\begin{align}
&\Phi_P^{(0)}(x)= -2\sqrt{\pi}S(x)\gamma^5\sin(Px/2),\label{phip0}\\
&\Phi_P^{(1)}(x)= \frac{\mu}{2\sqrt{\pi}}e^{\gamma_E}e^{ig^2\beta(x)}e^{-i\theta\gamma^5}\gamma^5\cos(P x/2)\; ,\label{phip1}
\end{align}
\end{subequations}
 
In the present paper we are interested in the effect of Lorentz contraction of the equal-time BS amplitude for which the separation of `quarks' is always spacelike. Therefore, instead of~(\ref{beta}) we use for the function $\beta(t,r)$ the simplified form:
\begin{equation}
\beta(0,r)=\frac{i}{2g^2}\left[\gamma_E+\ln\frac{\mu|r|}{2} +K_0(\mu|r|)\right].
\label{b0r}
\end{equation}
 
\subsection{ CM frame}
\label{CM}

Now let us analyze~(\ref{phipa}) in the center-of-mass frame. The two-momentum $P$ reduces here to $P=[\mu,0]$. The Minkowski product $P  x$ for the ET function equals zero, and due to the presence of $\sin(P  x/2)$ in~(\ref{phip0}) the amplitude in the sector $\Delta N=0$ vanishes identically. It is an interesting observation: in the center-of-mass frame the whole contribution for the ET BS wave-function comes only from nontrivial topological sectors and is given by
\begin{equation}
\Phi_P(0,r)=\sqrt{\frac{\mu}{2\pi |r|}}e^{\frac{1}{2}\gamma_E}e^{-i\theta\gamma^5}\gamma^5e^{-\frac{1}{2} K_0(\mu|r|)}.
\label{por}
\end{equation}

This result shows immediately that the simple contraction of the ET amplitude in each separate sector cannot take place, even in approximation! In any moving frame the two-momentum $P$ acquires a spatial component and the argument of the sine function  in~(\ref{phip0}) is no longer zero. For the moving `meson' $Px=P^1r\neq 0$, since we again consider ET function, naturally with respect to the new time. Consequently the simultaneous amplitude cannot be just the contracted CM ET function, which turned out to vanish identically. The boosting of a bound state in topologically nontrivial theory as Schwinger Model leads then to the mixing of various topological sectors. This conclusion is supported by our further results and can potentially apply to QCD bound states, i.e. to colliding hadrons, as well.

The object $\Phi_P$ is the complex matrix function (in spinor space) and therefore cannot be directly plotted. For this goal we need a real scalar function. Similarly as we did in our previous works~\cite{trrel,trren}, we define two gauge invariant quantities:
\begin{subequations}
\label{phip01}
\begin{align}
|\Phi_P^{(0)}(0,r)|&= \left(\frac{1}{2}\ \mathrm{tr}[\Phi_P^{(0)+}(0,r)\Phi_P^{(0)}(0,r)]\right)^{1/2},
\label{eq:modp0}\\
|\Phi_P^{(1)}(0,r)|&= \left(\frac{1}{2}\ \mathrm{tr}[\Phi_P^{(1)+}(0,r)\Phi_P^{(1)}(0,r)]\right)^{1/2},
\label{eq:modp1}
\end{align}
\end{subequations}
which are scalars. The first one obviously equals zero in the CM frame, and for the second we have
\begin{equation}
|\Phi_P^{(1)}(0,r)|=\sqrt{\frac{\mu}{2\pi |r|} }\  e^{\frac{1}{2} \gamma_E-\frac{1}{2}K_0(\mu |r|)}.
\label{nno}
\end{equation}

\begin{figure}[h]
\centering
{\includegraphics[width=0.48\textwidth]{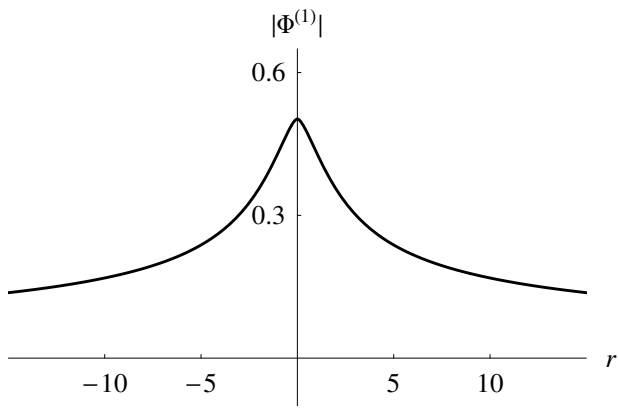}
\caption{The value of $|\Phi_{CM}^{(1)}|=|\Phi_{CM}|$  in units of $\mu$ as a function of relative distance between `quarks' in $\mu^{-1}$.}
\label{pl0}}
\end{figure}

The total strength of the BS amplitude may be defined as
\begin{align}
|\Phi_P(0,r)|&= \sqrt{\frac{1}{2}\ \mathrm{tr}[\Phi_P^{+}(0,r)\Phi_P(0,r)]}\nonumber\\
&=\sqrt{|\Phi_P^{(0)}(0,r)|^2+|\Phi_P^{(1)}(0,r)|^2}.
\label{topi}
\end{align}
In CM frame it is simply equal to~(\ref{nno}) and is plotted on Figure~\ref{pl0} as a function of spatial relative distance. Its behavior for large and small separations may be easily obtained from the known approximations~\cite{gr}:
\begin{equation}
K_0(x)\approx -\ln\frac{x}{2}-\gamma_E
\label{K0small}
\end{equation}
for $x\ll 1$, and
\begin{equation}
K_0(x)\approx \sqrt{\frac{\pi}{2x}}e^{-x}
\label{K0large}
\end{equation}
for $x\gg 1$. As may be easily verified, the value at maximum (i.e. for $r=0$) is equal to $\frac{\mu }{2\sqrt{\pi}} e^{\gamma_E}\approx 0.502\mu$ and for $|r|\rightarrow\infty$ we obtain
\begin{equation}
|\Phi_P(0,r)|\approx \sqrt{\frac{\mu}{2\pi |r|}}e^{\frac{1}{2}\gamma_E} \left(1-\frac{1}{2}\sqrt{\frac{\pi}{2\mu |r|}}e^{-\mu |r|}\right).
\label{asycm}
\end{equation}
The slow decay at large separations as $|r|^{-1/2}$ is a consequence of masslessness of quarks in this model. The exponential correction contains the effect of one massive meson created due to a quantum fluctuation~\cite{trrel}.

To distinguish the center-of-mass amplitudes from the ones in boosted frame, henceforth we will use the symbol $\Phi_{CM}(0,r)$ instead of $\Phi_P(0,r)$ for the former. 

In order to verify the Lorentz contraction, we should now compare the transformed quantity~(\ref{nno}) to the function $|\Phi_P(0,r)|$ found directly in a boosted frame.
If we had to do with the exact Lorentz contraction of the ET function, we would expect the result in the form 
$\sqrt{\gamma}|\Phi_{CM}(0,\gamma r)|$. Admittedly, the BS wave-function is not a probability amplitude, but the  strength $|\Phi_P(0,r)|$, we introduced, should be related (proportional) to it. It is then natural to expect from~(\ref{phip01}) and (\ref{topi}) similar transformation properties as those that apply to the true probability amplitude and hence the additional factor $\sqrt{\gamma}$ should appear~\cite{bettini}. It finds the confirmation in our numerical results. Let us for instance imagine one-dimensional world and a certain particle confined to a `box' of length L. The probability density for finding it at a given point $x$ is given by $P(x)$ subject to the obvious normalization condition:
\begin{equation}
\int\limits_0^L P(x)dx=1.
\label{pnorm}
\end{equation} 

Now let us assume that the system moves with velocity $v$ with respect to certain observer. The normalization of probability density in the observer's frame (i.e. $P'(x)$) must be preserved, but due to the Lorentz contraction instead of $L$ the box has now length $L/\gamma$. Consequently 
\begin{equation}
\int\limits_0^{L/\gamma} P'(x)dx=1.
\label{pnorma}
\end{equation}
If one substitutes for $P'(x)$ the modified density:
\begin{equation}
P'(x)=\gamma P(\gamma x),
\label{ppm}
\end{equation}
it may be easily verified that the condition~(\ref{pnorma}) is automatically satisfied. Of course this simple argument has not to be applied to a box, but equally well to any interval $dx$ and it is a simple consequence of the transformation of the integration measure. Alternatively one can look on $P(x,t)$ as a temporal component of the probability current.
 
\subsection{Boosted frame}
\label{boost} 

Now we need the ET function in the frame in which the meson moves with momentum $p$ (laboratory frame). 
Surely, we have $P=[E_p, p]=[\sqrt{\mu^2+p^2}, p]$. The ET functions~(\ref{phip01}), but now with respect to the new time, become:
\begin{subequations}
\label{psib}
\begin{align}
|\Phi_P^{(0)}(0,r)|&=\sqrt{\frac{\mu}{2\pi |r|}}e^{\frac{1}{2}\gamma_E} e^{\frac{1}{2} K_0(\mu|r|)}\left|\sin\frac{pr}{2}\right|,\label{psib0}\\
|\Phi_P^{(1)}(0,r)|&=\sqrt{\frac{\mu}{2\pi |r|}}e^{\frac{1}{2}\gamma_E} e^{-\frac{1}{2} K_0(\mu|r|)}\left|\cos\frac{pr}{2}\right|.
\label{psib1}
\end{align}
\end{subequations}

Contrary to~(\ref{nno}) both expressions $|\Phi_P^{(0)}(0,r)|$ and $|\Phi_P^{(1)}(0,r)|$ -- apart from decreasing factors -- have now the oscillating character.
The presence of sine and cosine with arguments $p r/2$ in the above formulas makes the independent Lorentz contraction of $\Phi_P^{(0)}$ and $\Phi_P^{(1)}$  impossible to happen. It is clear than none of these amplitudes can be Lorentz contracted CM function~(\ref{por}), but, as we show below, it still occurs for $|\Phi_P|$.

Consider first the limits of small and large distances. For $r\rightarrow 0$ we get:
\begin{subequations}
\label{asy}
\begin{align}
|\Phi^{(0)}_P(0,r)|&\approx \frac{1}{\sqrt{\pi} |r|}\left|\sin\frac{p r}{2}\right|,\label{asy0}\\
|\Phi^{(1)}_P(0,r)|&\approx \frac{\mu}{2\sqrt{\pi}}e^{\gamma_E}\left|\cos\frac{p r}{2}\right|,\label{asy1}
\end{align}
\end{subequations}
and for the full amplitude we have:
\begin{align}
|\Phi_P(0,r)|&\approx\frac{1}{\sqrt{\pi}}\left(\frac{1}{r^2}\sin^2\frac{pr}{2}+\frac{\mu^2}{4}e^{2\gamma_E}\cos^2\frac{pr}{2}\right)^{1/2}\nonumber\\
&\underset{r\rightarrow 0}{\longrightarrow}\frac{\mu}{2\sqrt{\pi}}e^{\gamma_E}\sqrt{1 +\frac{p^2}{\mu^2}e^{-2\gamma_E}}\nonumber\\
&=|\Phi_{CM}(0,\gamma\cdot 0)|\sqrt{1 +\frac{p^2}{\mu^2}e^{-2\gamma_E}}.
\label{full0}
\end{align}
Let us recall that for the exact Lorentz contraction we would get the result:
\begin{equation}
\sqrt{\gamma} |\Phi_{CM}(0,\gamma\cdot 0)|=\left(1 +\frac{p^2}{\mu^2}\right)^{1/4}|\Phi_{CM}(0,\gamma\cdot 0)|,
\label{compa}
\end{equation}
since $\gamma=\frac{E_p}{\mu}=\sqrt{1+\frac{p^2}{\mu^2}}$. This deviation is, however, acceptable. For instance for $\gamma=1.5$ or $p^2=1.25\mu^2$ the true coefficient would be equal to $\sqrt{\gamma}\approx 1.22$, whilst for the factor in~(\ref{full0}) we find:
$$
\sqrt{1 +\frac{p^2}{\mu^2}e^{-2\gamma_E}}\approx 1.18,
$$
so the ratio
\begin{equation}
\kappa=\frac{\left(1+\frac{p^2}{\mu^2}e^{-2\gamma_E}\right)^{1/2}}{\left(1+\frac{p^2}{\mu^2}\right)^{1/4}}
\label{kappa}
\end{equation}
gives $0.97$ in this case. Similarly, for $\gamma=3$ one gets $\kappa\approx 1.09$ and for $\gamma=6$, $\kappa\approx 1.43$. These results are reflected on the exact plots presented in the following section.

Now let us turn to the asymptotic behavior for large separations $r$. Using~(\ref{K0large}), we obtain:
\begin{subequations}
\label{asyinf}
\begin{align}
|\Phi^{(0)}_P(0,r)|\approx &\sqrt{\frac{\mu}{2\pi |r|}}e^{\gamma_E/2}\left|\sin\frac{p r}{2}\right|\nonumber\\ & \times\left(1+\frac{1}{2}\sqrt{\frac{\pi}{2\mu |r|}}e^{-\mu |r|}\right),\label{asyinf0}\\
|\Phi^{(1)}_P(0,r)|\approx &\sqrt{\frac{\mu}{2\pi |r|}}e^{\gamma_E/2}\left|\cos\frac{p r}{2}\right|\nonumber\\ & \times\left(1-\frac{1}{2}\sqrt{\frac{\pi}{2\mu |r|}}e^{-\mu |r|}\right).\label{asyinf1}
\end{align}
\end{subequations}
Consequently the full amplitude for $|r|\gg \mu^{-1}$ may be written in the form:
\begin{equation}
|\Phi_P(0,r)|\approx \sqrt{\frac{\mu}{2\pi |r|}}\ e^{\frac{1}{2}\gamma_E}\left(1- \frac{\pi}{2\mu |r|} \ e^{-\mu|r|} \cos pr\right)^{1/2}.
\label{fullinf}
\end{equation}
As can be easily seen, at least asymptotically, where highly energetic quantum fluctuations may be ignored and the second term under the square root in~(\ref{fullinf}) is negligible, the Lorentz contraction in the form expected by relativistic kinematics takes place, since one has
\begin{equation}
|\Phi_P(0,r)|\approx \sqrt{\frac{\mu}{2\pi |r|}}e^{\gamma_E/2}\approx \sqrt{\gamma}\,|\Phi_{CM}(0,\gamma r)|
\label{ascont}
\end{equation}
(see the discussion in the following section).
The deviations from the exact Lorentz contraction of the full function are then exponentially small. It should be again stressed, that we owe this simple result to the taking into account all contributing instanton sectors.  

At the end of this section let us consider the exact formulas~(\ref{psib}). If we calculate $|\Phi_P(0,r)|$ according to~(\ref{topi}) and~(\ref{psib}), we find:

\begin{align}
|\Phi_P(0,r)|=&\sqrt{\frac{\mu}{2\pi |r|}}e^{\frac{1}{2}\gamma_E}\bigg[e^{K_0(\mu |r|)}\sin^2\frac{p r}{2}\nonumber\\
&+e^{-K_0(\mu |r|)}\cos^2\frac{p r}{2}\bigg]^{1/2}.\nonumber\\
\label{fufa}
\end{align} 
It is obvious, that without exponential factors (in~(\ref{fufa}) and in~(\ref{nno})) one might make use of the Pythagorean trigonometric identity for the terms in square brackets and the phenomenon would be exact. One sees then in a clear way, how contributions from all instanton sectors complement one another and are all necessary to satisfy the relativistic transformation property. 

The expression~(\ref{fufa}) may be compared with Lorentz-contracted ET function from the CM frame. We get
\begin{align}
\frac{|\Phi_P(0,r)|}{\sqrt{\gamma}|\Phi_{CM}(0, r)|}=&\left[\cos^2\frac{p r}{2}+e^{2K_0(\mu |r|)}\sin^2\frac{p r}{2}\right]^{1/2}.
\label{com}
\end{align}

When $K_0(\mu  |r|)$ approaches zero, the ratio becomes unity. As we have seen, this happens for large distances. One may then say that amplitude `tails' respect Lorentz contraction. Since $K_0(x)$ is a monotonically descending function,  we see, that when $|r|$ decreases, the coefficient $e^{2K_0(\mu  |r|)}$ becomes larger and the deviation from the Lorentz contractions becomes more significant.

There are also particular spatial points for which the exact contraction occurs. In the laboratory frame, where the meson moves with momentum $p$, they are characterized by conditions $p |r|=2 n\pi$ where $n=1,2,3,\ldots$. At these points the sine function vanishes, and~(\ref{com}) equals $1$. The whole contribution to BS amplitude comes here from instanton sectors $\Delta N=\pm 1$. The $r$-dependence of the ratio of contraction -- although in another layout -- has been observed in~\cite{diet,hoyer3}.

\section{Numerical results and conclusions}
\label{conclu} 

\begin{figure}[t]
\centering
{\includegraphics[width=0.48\textwidth]{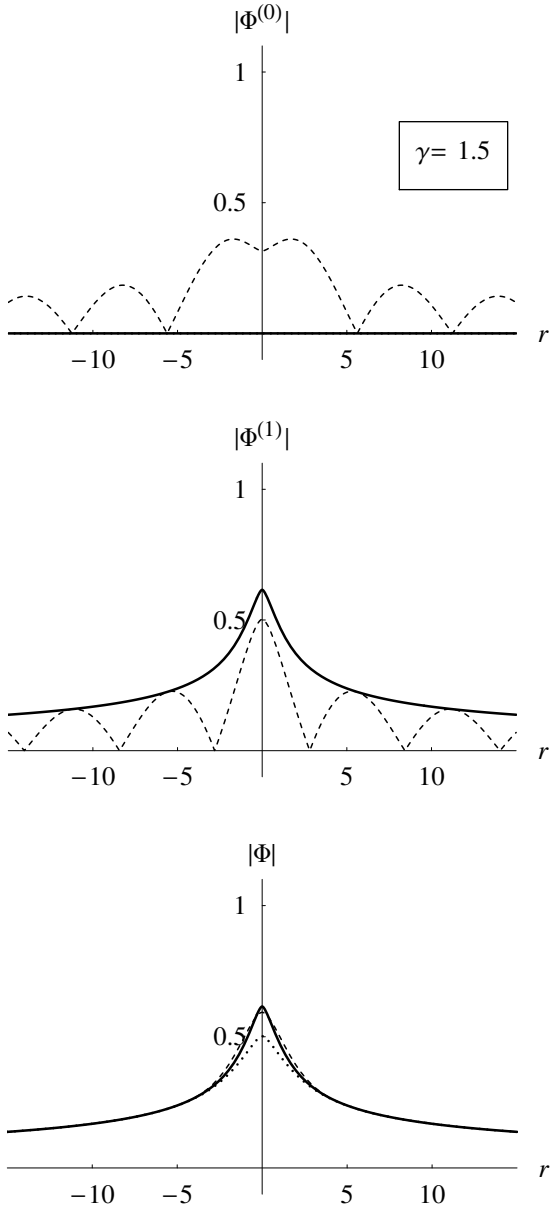}
\caption{Spatial distribution of $|\Phi^{(0)}_P|$ (upper plot), $|\Phi^{(1)}_P|$ (central plot) and $|\Phi_P|$ (lower plot) in units of $\mu$ (dashed lines) and corresponding boosted ET wave-functions from the CM frame (solid lines). On the upper plot the solid line is not visible since it overlaps with the horizontal axis. The dotted curve represents the  uncontracted function. The distance $r$ is measured in $\mu^{-1}$. The value of $\gamma$ is $1.5$.}
\label{pl1}}
\end{figure}

In this section we would like to summarize the obtained results and present plots revealing the exact behavior of BS amplitudes. In Figure~\ref{pl1} the plots of the spatial distributions of $|\Phi^{(0)}_P(0,r)|$, $|\Phi^{(1)}_P(0,r)|$ and $|\Phi_P(0,r)|$ are presented (as dashed lines) and compared with the boosted ET functions $\sqrt{\gamma}|\Phi^{(0)}_{CM}(0,\gamma r)|$, $\sqrt{\gamma}|\Phi^{(1)}_{CM}(0,\gamma r)|$ and $\sqrt{\gamma}|\Phi_{CM}(0,\gamma r)|$ (solid lines) for relatively small value of the parameter $\gamma=1.5$. As we know from the previous section, the ET amplitude for the instanton sector $\Delta N=0$ vanishes identically, so the solid line on the upper plot coincides with $r$ axis.

The oscillating character of the exact amplitudes in the laboratory frame (where the meson is moving) is clearly visible. No simple Lorentz contraction is then observed for contributions from separate topological sectors. On the other hand the agreement between $|\Phi_P(0,r)|$ and $\sqrt{\gamma}|\Phi_{CM}(0,\gamma r)|$ is very good.

The dotted curve on the lower plot represents the uncontracted function, as if Galilean and not Einsteinian kinematics were legitimate. This is in fact the graph of figure~\ref{pl0}, redrawn here for comparison. 
The fact that this curve agrees with the other two, results from the very special asymptotic behavior of the amplitude, which in SM decreases as $|r|^{-1/2}$. Consequently the two $\gamma$-dependent factors cancel each other in the expression $\sqrt{\gamma}|\Phi_{CM}(0, \gamma r)|$.
It is obvious that with the dashed line representing the true behavior one would obtain distributions of various physical quantities more squeezed than with dotted one.

\begin{figure}[h]
\centering
{\includegraphics[width=0.48\textwidth]{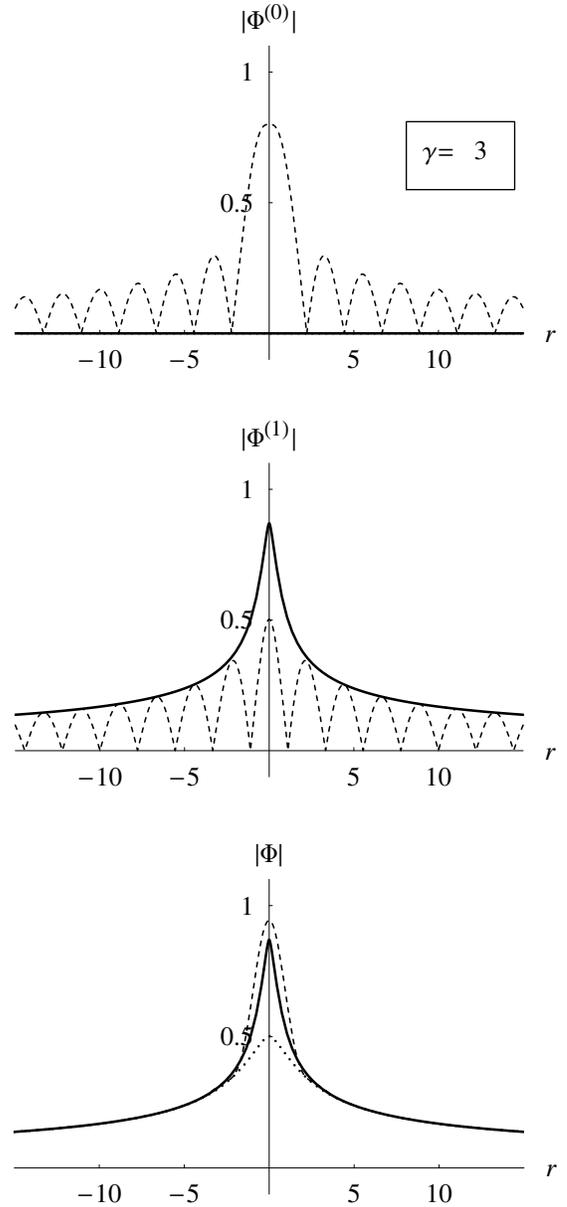}
\caption{Same as Figure~\ref{pl1} but for $\gamma=3$.} 
\label{pl2}}
\end{figure}

In Figure~\ref{pl2} similar results are presented but for moderate value of $\gamma=3$. For faster moving meson the period of oscillations is smaller, but the general properties of BS amplitudes are preserved. As we found in formula~(\ref{full0}) the discrepancy between both amplitudes becomes more significant for smaller values of the relative distance.  If $p^2$ is large then the ones in the numerator and denominators of the ratio $\kappa$ between exact function at $r=0$ and the contracted ET CM function may be neglected. We have then the expression
\begin{equation}
\kappa\approx \sqrt{\frac{|p|}{\mu}}\ e^{-\gamma_E}\approx 0.56\ \sqrt{\frac{|p|}{\mu}},
\label{fact}
\end{equation}
which increases with rising momentum. 

\begin{figure}
\centering
{\includegraphics[width=0.48\textwidth]{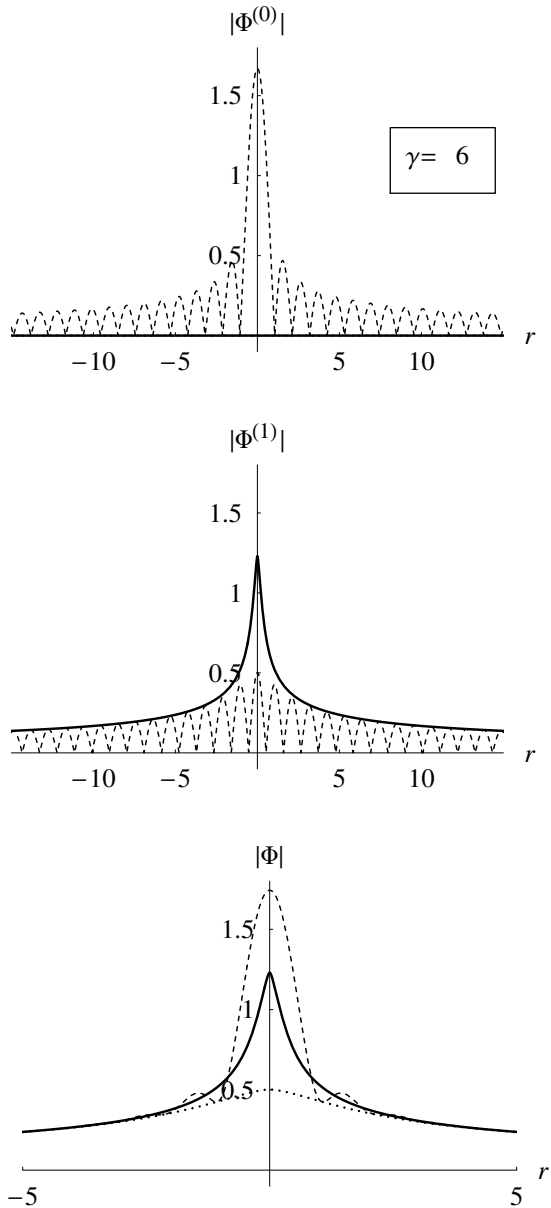}
\caption{Same as Figure~\ref{pl1} but for $\gamma=6$. Please, note the change of scale on the horizontal axis in the last plot.} 
\label{pl3}}
\end{figure}

The principal properties observed for small and moderate values of $\gamma$ manifest themselves even more strongly for $\gamma=6$. The region of deviation from the exact Lorentz contraction is becoming narrower with increasing meson velocity (one should note the change of scale in the last plot of Figure~\ref{pl3}). 

These results may be compared with the deformation of pion amplitudes in Gross-Neveu~\cite{gn} model with increasing pion momentum as shown in Figure~7 of~\cite{st}. 

Summarizing one can say, that Lorentz contraction is of course not exact but relatively well satisfied in general. This conclusion stays with agreement with previous results obtained within certain approximations. The true transformations are reflections of nontrivial dynamical character of boosts expressed in commutator:
\begin{equation}
[K,H]=iP\neq 0.
\label{hk}
\end{equation} 
It would be hard to imagine, that all Fock sectors deform identically under boosts, so this result was expected.

The deviations from the Lorentz contraction appear for small distances between quarks, but slowly decaying `tails' always respect the phenomenon. This effect is understandable, since at small distances greater role is played by higher sectors, connected with short-lived quantum fluctuations, especially because the Schwinger boson is a massive particle.

This analysis seems to be of certain importance, since it is based on the {\em exact} BS amplitudes and their behavior may be explicitly demonstrated. The underlying dynamics of quantum fields stays in agreement with relativistic kinematics. The knowledge of the behavior of the amplitudes deformations under boosts is essential, because these functions are generally only known in the CM systems, while in collisions we are dealing with highly accelerated particles or nuclei.
 
The observation that Lorentz contraction requires the inclusion of all contributing instanton sectors seems to support the Nakanishi's suggestion, that topological vacua are not Lorentz invariant. This point deserves a detailed explanation in the future, since it may seem counterintuitive.

\section*{Acknowledgments}
The author would like to thank to Professors Iwo Bia{\l}ynicki-Birula and Krzysztof Meissner for their suggestions.


\begin{thebibliography}{99}
\bibitem{baltz} A. Baltz {\em et al.}, Phys. Rept. {\bf 458}, 1(2008).
\bibitem{singh} R. Singh {em et al.}, Adv. High Energy Phys. {\bf 2013}, 761474(2013). 
\bibitem{wong} C.-Y. Wong, {\em Introduction to High-energy Heavy-ion Collisions}, World Scientific, London 1994.
\bibitem{bro1} S.J. Brodsky and J.R. Primack, Ann. Phys. {\bf 52}, 315(1969).
\bibitem{bro2} S.J. Brodsky, {\em Brandei Lectures 1969, vol. 1}, Gordon and Breach, New York 1971.
\bibitem{kimza} Y.S. Kim and R. Zaoui, Phys. Rev. {\bf D4}, 1764(1971).
\bibitem{inozem} V.I. Inozemtsov, preprint P2-10044, Dubna 1976.
\bibitem{kimnoz} Y.S. Kim, M.E. Noz and S.H. Oh, Found. Phys. {\bf 9}, 947(1979).
\bibitem{hoyer1} P. Hoyer, Phys. Lett {\bf B172}, 101(1986).
\bibitem{gnf} W. Gl\"ockle, Y. Nogami and I. Fukui, Phys. Rev. {\bf D 5}, 584(1987).
\bibitem{gn}  W. Gl\"ockle and Y. Nogami, Phys. Rev. {\bf D35}, 3840(1987).
\bibitem{kim} Y.S. Kim, in `Dartmouth 1997, Theoretical physics', 269(1997).
\bibitem{jarv0} M. J\"arvinen, Phys. Rev. {\bf D70}, 065014(2004), {\em ibid.} {\bf D 71}, 085006(2005).
\bibitem{hoyer2} P. Hoyer, AIP Conf. Proc. {\bf 904}, 65(2007).
\bibitem{diet}  D.D. Dietrich, P. Hoyer and M. J\"arvinen, Phys. Rev. {\bf D85}, 105016(2012).
\bibitem{hoyer3} P. Hoyer, arXiv:1402.5005.
\bibitem{simonov} Yu.A. Simonov, Phys. Rev.  {\bf D91}, 065001(2015).
\bibitem{bs} E. E. Salpeter and H.A. Bethe, Phys. Rev. {\bf 84}, 1232(1951).
\bibitem{gml} M. Gell-Mann and F. Low, Phys. Rev. {\bf 84}, 350(1951).
\bibitem{wick} G. C. Wick, {Phys. Rev.} {\bf 96}, 1124(1954).
\bibitem{cut} R. E. Cutkosky, Phys. Rev. {\bf 96}, 1135(1954).
\bibitem{nn1} N. Nakanishi, Prog. Theor. Phys. Suppl. {\bf 95}, 1(1988).
\bibitem{baka} R. Bakamjian and L.H. Thomas, Rhys. Rev. {\bf 92}, 1300(1953).
\bibitem{js} J. Schwinger, in {\em Theoretical Physics}, Trieste Lectures 1962 (I.A.E.A. Vienna 1963), p. 89; Phys. Rev. {\bf 128}, 2425(1962).
\bibitem{trsing} T. Rado\.zycki, Phys. Rev. {\bf D75}, 085005(2007).
\bibitem{trfactor} T. Rado\.zycki, Acta Phys. Polon. {\bf B40}, 1653(2009).
\bibitem{abh} C. Adam, R.A. Bertlmann and P. Hofer , Riv. Nuovo Cim. {\bf 16}, 1(1993).
\bibitem{cadam1} C. Adam, Z. Phys. {\bf C63}, 169(1994).
\bibitem{smil} A. V. Smilga, Phys. Rev. {\bf D49}, 5480(1994).
\bibitem{gattr} C. Gattringer, hep-th/9503137.
\bibitem{maie} G. Maiella and F. Schaposnik, Nucl. Phys. {\bf B132}, 
357(1978).
\bibitem{rot} K. D. Rothe and J. A. Swieca, Ann. Phys. {\bf 117}, 382(1979).
\bibitem{gmc} G. McCartor, Int. J. Mod. Phys. {\bf A12}, 1091(1997).
\bibitem{rajar} R. Rajaraman, {\em An Introductions to Solitons and Instantons in Quantum Field Theory}, North-Holland, New York 1982.
\bibitem{huang} K. Huang, {\em Quarks, Leptons and Gauge Fields}, 
World Scientific, London 1992. 
\bibitem{cdg} C.G. Callan, R.F. Dashen and D.J. Gross, Phys. Lett. {\bf 
63B}, 334(1976).
\bibitem{trinst} T. Rado\.zycki, Phys. Rev. {\bf D 60}, 105027(1999).
\bibitem{eden} R. J. Eden, Proc. Roy. Soc. {\bf A 217}, 390(1953).
\bibitem{mand} S. Mandelstam, Proc. Roy. Soc. Lond. {\bf A 233}, 248(1955).
\bibitem{trjmn} T. Rado\.zycki and J.M. Namys{\l}owski, Phys. Rev.
{\bf D 59}, 065010(1999).
\bibitem{naka} N. Nakanishi, Prog. Theor. Phys. {\bf 58}, 1927(1977).
\bibitem{abe} M. Abe and S. Miyake, Prog. Theor. Phys. {\bf 59}, 2037(1978).
\bibitem{trrel} T. Rado\.zycki, Phys. Rev. {\bf D87}, 085042(2013).
\bibitem{trren} T. Rado\.zycki, Eur. Phys. J.  {\bf C74}, 3037(2014).
\bibitem{gr} For instance I. S. Gradshteyn's and I. M. Ryzhik's, {\em Table of Integrals
Series and Products}, Academic Press, New York 1980.
\bibitem{bettini} A. Bettini, {\em Introduction to Elementary Particle Physics}, Cambridge University Press, New York 2012.
\bibitem{gn} D.J. Gross and A. Neveu, Phys. Rev. {\bf D 10}, 3235(1974).
\bibitem{st} V. Sch\"on and M. Thies, hep-th/0008175.
\end{thebibliography}
\end{document}